\documentclass{PoS}
\usepackage{amsmath,amssymb,graphicx,multirow,tabularx,physics}
\usepackage{davitex}
\usepackage{graphicx}
\usepackage{adjustbox}
\usepackage{tikz,pgfplots}
\usepackage{float}
\usetikzlibrary{shapes.geometric, arrows}

\title{Euclidean correlation functions of the topological charge density}

\ShortTitle{Euclidean correlation functions of the topological charge density}

\author{\speaker{Lukas Mazur}\\
Fakult\"at f\"ur Physik, Universit\"at Bielefeld, D-33615
Bielefeld, Germany\\ 
E-mail: \email{lmazur@physik.uni-bielefeld.de}}

\author{Luis Altenkort\\
Fakult\"at f\"ur Physik, Universit\"at Bielefeld, D-33615
Bielefeld, Germany\\
E-mail: \email{altenkort@physik.uni-bielefeld.de}}

\author{Olaf Kaczmarek\\
Fakult\"at f\"ur Physik, Universit\"at Bielefeld, D-33615
Bielefeld, Germany;\\
Key Laboratory of Quark \& Lepton Physics (MOE) and Institute of
Particle Physics, \\Central China Normal University, Wuhan 430079, 
China\\
E-mail: \email{okacz@physik.uni-bielefeld.de}}

\author{Hai-Tao Shu\\
Fakult\"at f\"ur Physik, Universit\"at Bielefeld, D-33615
Bielefeld, Germany\\
E-mail: \email{htshu@physik.uni-bielefeld.de}}

\abstract{
    We present first results of our study on the Euclidean topological charge density correlation function.
    In order to get a well defined topological charge density and to improve the signal of the correlation function at large separations we make use of the gradient flow.
    We investigate the flow-time dependence on fine quenched lattices. The final goal of this study is to perform a continuum extrapolation for the pure SU(3) plasma and to extract the related transport coefficient, the sphaleron rate. 
}

\FullConference{37th International Symposium on Lattice Field Theory - 
Lattice2019\\
16-22 June 2019\\
Wuhan, China}

\begin{document} \linespread{0.83}
\section{Introduction}
The change of the topological charge is given by the sphaleron rate.
It is a transport coefficient that is important for the lifespan of any chiral quark number in the quark-gluon-plasma \cite{Moore:2010jd,Giudice:1993bb}.
Perturbative calculations of the sphaleron rate exist \cite{Moore:2010jd,Giudice:1993bb}, but they are not reliable in the relevant coupling region for heavy ion phenomenology. 
For that reason, non-perturbative calculations are required from lattice QCD.
However, on the lattice the gluonic definition of topological charge density is affected by UV-fluctuations. This leads to a poor signal-to-noise ratio which prevents us from high precision estimations of the two-point correlation functions, especially at large separations. To overcome this difficulty we adopt the gradient flow technique \cite{Luscher:2010iy,Luscher:2011bx,Narayanan:2006rf}. The field smearing nature of the gradient flow, i.e. the smoothing of the gauge fields, allows a well-defined topological charge \cite{Luscher:2010iy,Luscher:2010we} and a good signal of the correlation function. This provides us access to investigate how the instantons emerge in the continuum limit of lattice QCD. Some related work can be found in \cite{Ce:2015qha,Taniguchi:2016tjc,Kotov2018}.
In this work we analyze the gradient-flowed topological charge density correlation function, that is required for the computation of the sphaleron rate on the lattice.

\section{The correlation function and the gradient flow}\label{sec:theory}
On the lattice the sphaleron rate can be obtained from the zero frequency limit of the spectral function, $\Gamma_{sphal}=\lim_{\omega\rightarrow0}\frac{2T\rho_{q}(\omega,0)}{\omega}$, of its Euclidean correlation function
\begin{equation}
    G_{q}(\tau,\vec{p})=\int d^{3}x\,\,\, e^{i\vec{p}\cdot\vec{x}}\langle q(\vec{x},\tau)q(\vec{x},0)\rangle=\int_{0}^{\infty}\frac{d\omega}{\pi}\rho_{q}(\omega,\vec{p})\frac{\cosh\omega\left(\frac{1}{2T}-\tau\right)}{\sinh\frac{\omega}{2T}},
\end{equation}
where $q(\vec{x},\tau)$ is the topological charge density. The lattice correlation function for $\vec{p}=0$ can be calculated from:
\begin{equation}
    G^{\text{lat}}_{q}(\tau)=\frac{1}{V}\sum_{\vec{x}} \langle q(\vec{x},0)q(\vec{x},\tau) \rangle,
\end{equation}
where the topological charge density is defined by: 
\begin{equation}\label{eq:topCh_den}
    q(x)=\frac{g^2}{32\pi^{2}}\epsilon_{\mu\nu\rho\sigma}\textrm{tr}\left\{ F_{\mu\nu}(x)F_{\rho\sigma}(x)\right\}.
\end{equation}
On the lattice we use a highly-improved field-strength tensor $F_{\mu\nu}$ \cite{BilsonThompson:2002jk}.
The topological charge density $q(x)$ is anti-symmetric. Hence, due to arguments
based on reflection positivity \cite{Vicari:1999xx}, this correlator has to be negative for
all nonzero separations $|x| > 0$ in the continuum. Since it is also possible to construct the topological
susceptibility from integrating this correlator and because of the fact that
the susceptibility has to satisfy $\chi_{q} \geq 0$, it follows that the correlator has to have a
positive contact term at $x = 0$.

The topological charge density as defined in \eqref{eq:topCh_den} is affected by UV-fluctuations on the lattice. 
Therefore, we need to smoothen the configurations in order to obtain a well defined topological charge. Many smoothing techniques exist. In this work we are using the gradient flow.
The gradient flow introduces an extra coordinate $t$ (flow-time) and defines a $t$-dependent gauge field $B_{\mu}(x,t)$ \cite{Luscher:2010iy}
\begin{equation}
    \frac{d}{dt}B_{\mu}(x,t)  =D_{\nu}G_{\nu\mu}(x,t),\qquad D_{\mu}=\partial_{\mu}+\left[B_{\mu}(x,t),\,\,\mathord{\cdot}\,\,\right],
\end{equation}
\begin{equation}
    G_{\mu\nu}(x,t)=\partial_{\mu}B_{\nu}(x,t)-\partial_{\nu}B_{\mu}(x,t)+\left[B_{\mu}(x,t),B_{\nu}(x,t)\right]
\end{equation}
with the initial condition $B_{\mu}(x,t)|_{t=0}  =A_{\mu}(x)$.
$t$ has the dimension of inverse mass-squared. The flow smoothens the fields over a region of radius $r_{F}=\sqrt{8t}$.
We use a Symanzik improved gradient flow (Zeuthen flow \cite{Ramos:2015baa}) and integrate the flow equation by using a 3$^{\text{rd}}$ order Runge-Kutta method with an adaptive step-size algorithm.

\section{Setup and simulation parameters}
The SU(3) pure gauge configurations that are used in this work are listed in table \ref{table:conf}. They were generated using the quenched approximation at a temperature $T/T_c\approx1.5$.
All configurations are separated by 500 sweeps, where one sweep consists of one heat bath and four overrelaxation steps.
{\small  \begin{table}[t]
    \centering
\begin{tabular}{|c|c|c|c|c|}
    \hline
    $N_s$	&	  $N_{\tau}$	&	  $\beta$	&	 $a$ [fm]	&	 \#conf	\tabularnewline
    \hline\hline
	$64$	&	 $16$	&	 6.873	&	 0.026	&	 10000	\tabularnewline
	$80$	&	 $20$	&	 7.035	&	 0.022	&	 10000	\tabularnewline
	$96$	&	 $24$	&	 7.192	&	 0.018	&	 10000	\tabularnewline
	$120$	&	 $30$	&	 7.394	&	 0.014	&	 10000	\tabularnewline
    \hline
\end{tabular}

    \caption{Lattice dimensions, $\beta$-values, lattice spacings and number of configurations which are used in this work.}
    \label{table:conf}
\end{table}}
The lattice spacing $a$ has been determined by the Sommer parameter $r_0$ \cite{Sommer:1993ce}, where we use a parametrization from \cite{Francis:2015lha} with updated coefficients from \cite{Burnier:2017bod}.

\begin{figure}[t]
    \centering
\begin{minipage}{0.49\textwidth}
    \adjincludegraphics[width=1\textwidth,valign=t]{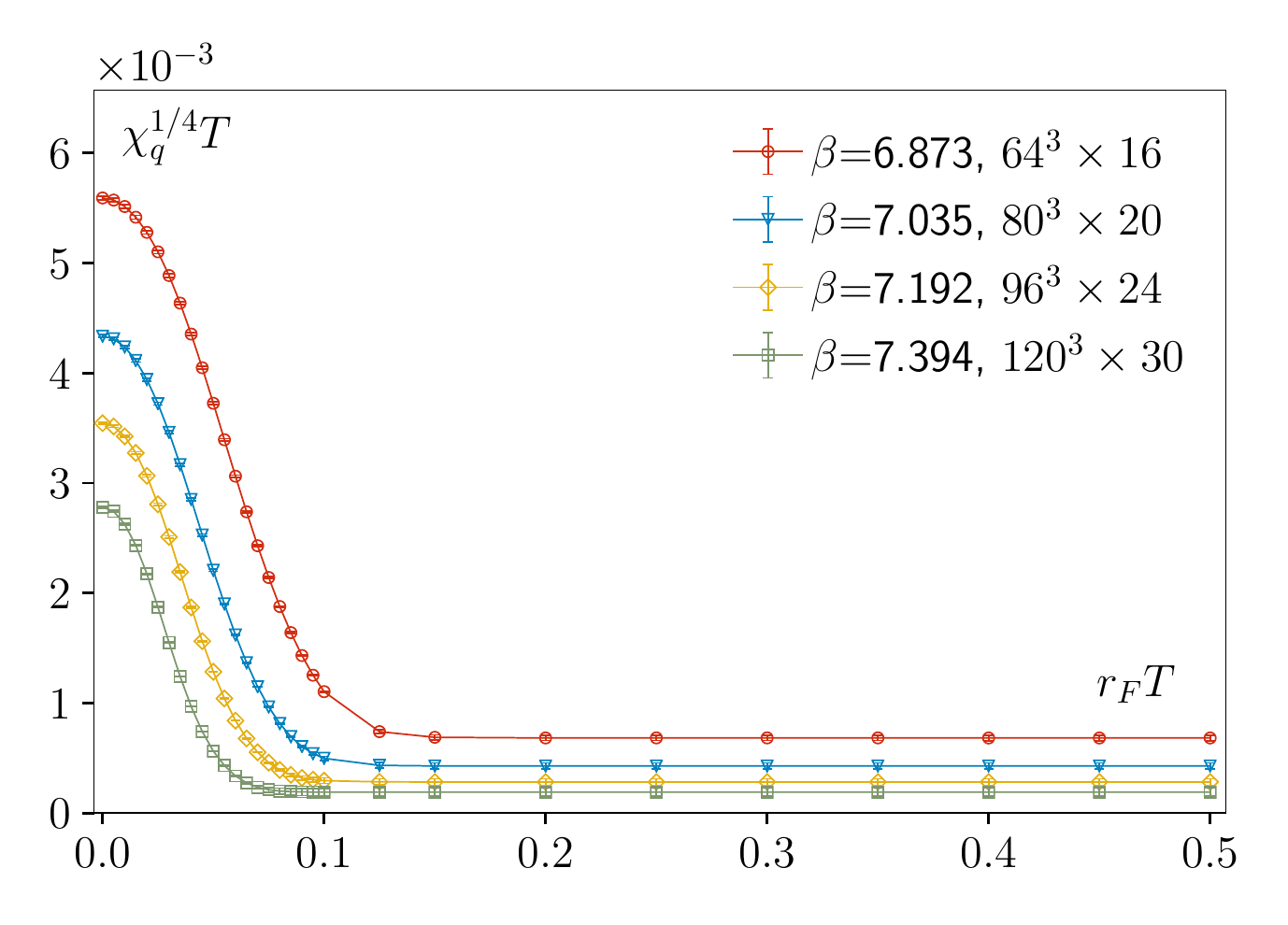}
\end{minipage}
\begin{minipage}{0.49\textwidth}
    \adjincludegraphics[width=1\linewidth,valign=t]{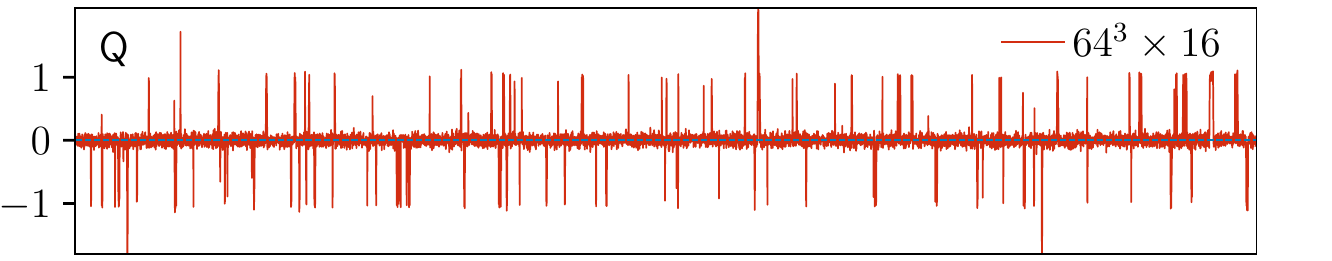}
    \adjincludegraphics[width=1\linewidth,valign=t]{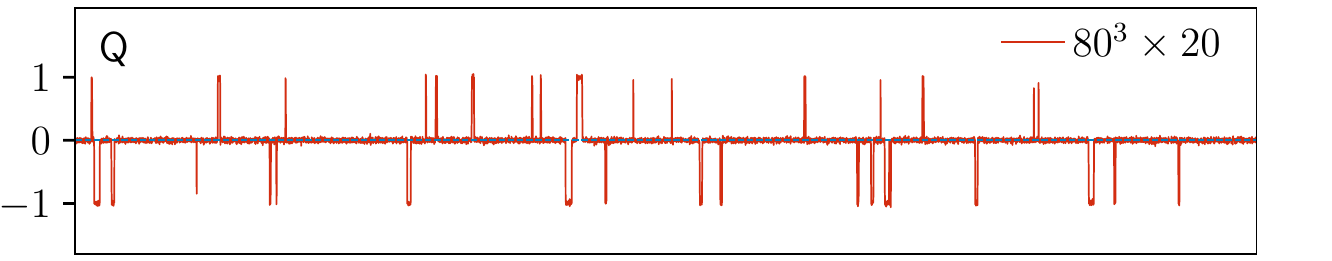}
    \adjincludegraphics[width=1\linewidth,valign=t]{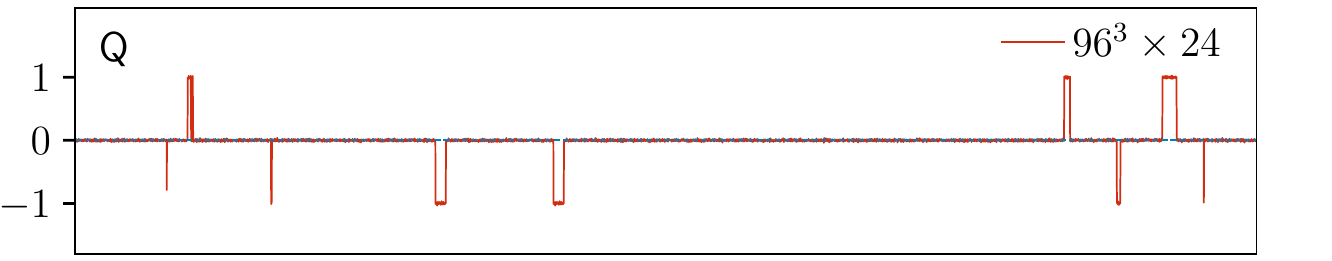}
    \adjincludegraphics[width=1\linewidth,valign=t]{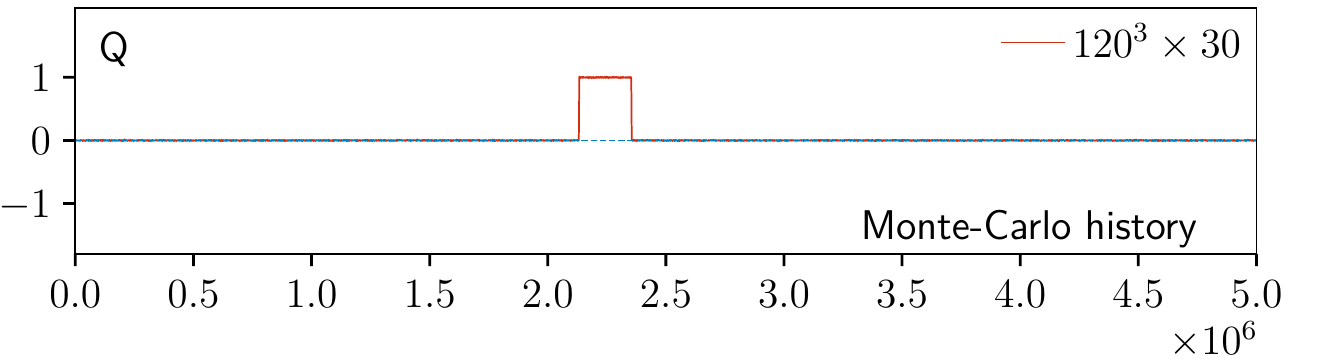}
\end{minipage}
    \caption{Left: The topological susceptibility with respect to the smoothing radius.
    Right: Trajectory of the topological charge at $r_{F}T=0.15$.}
    \label{fig:sus_all}
\end{figure}

\section{How much flow is necessary?}\label{sec:top_sus}

Since the gluonic definition of the topological charge density is only valid on smooth configurations, the same should also hold for the correlation function $G^{\text{lat}}_{q}(\tau)$. To figure out a lower bound for the flow-time $t$ at which the configurations are smooth enough, one can take a look at the topological susceptibility.

Integrating $q(x)$ yields the topological charge,
\begin{equation}
    Q=\int d^{4}x\,q(x),
\end{equation}
from which we can also compute the topological susceptibility: 
\begin{equation}
    \chi_{q}= \frac{ \langle Q^2 \rangle}{V} = \int d^4x \,\langle q(x)q(0)\rangle.
\end{equation}
The left hand side of figure \ref{fig:sus_all} shows a plot of the susceptibility with respect to the flow-time. 
In order to compare the effects of the flow across lattices with different lattice spacings we use the rescaled dimensionless smoothing radius $r_{F}T=\sqrt{8t}T$.
We observe that at small flow-times the susceptibility is still affected by UV-fluctuations. With increasing flow-time these fluctuations decrease. When all UV-fluctuations have been smoothed out, the susceptibility reaches a plateau. It is now natural to choose as a lower bound the flow-time at which the plateau starts.
One can see that the coarser lattices need more smoothing to reach the plateau than finer lattices. A good compromise is then to choose the lowest $r_{F}T$ at which the plateau starts from the coarsest lattice. In our case $r_{F}T=0.15$ seems to be a good choice.

\section{Is topological tunneling sufficient?}

When dealing with topology, one should check whether the topological charge freezes or not. On a periodic continuum gauge field the topological charge cannot change by any continuous deformation, while on a periodic lattice gauge field it can "tunnel" through non-continuum-like configurations. That amounts to having less topological fluctuations on lattices which are closer to continuum, i.e as the lattice spacing decreases, the topological freezing increases. That leads to large autocorrelation times.

In the plot of the right hand side of figure \ref{fig:sus_all} the trajectory of the topological charge of all lattices is shown. Indeed, we see that the fluctuations decrease as the lattice spacing decreases. However, we see that the $N_{\tau}=24$ lattice is still not entirely frozen, while for the finest lattice only one tunneling is observed.

\section{Which part of the correlation function is reliable?}
Having found the minimum flow-time which is needed to have a valid definition of $G^{\text{lat}}_{q}(\tau)$ in section \ref{sec:top_sus}, the next question is at which separations the correlation functions are still reliable, i.e. not affected by the flow. 
One may expect that the correlation is influenced by flow effects at separations which do not fulfill the condition 
\begin{equation}\label{eq:flow_limit}
2r_{F}T < \tau T.
\end{equation}
The reason is that if we use a smoothing radius larger than $\tau T / 2$, the smoothing radii of the two sources in the correlation function overlap. 

\begin{figure}[t]

\begin{minipage}{0.50\textwidth}
    \raggedleft
    \includegraphics[width=1\linewidth]{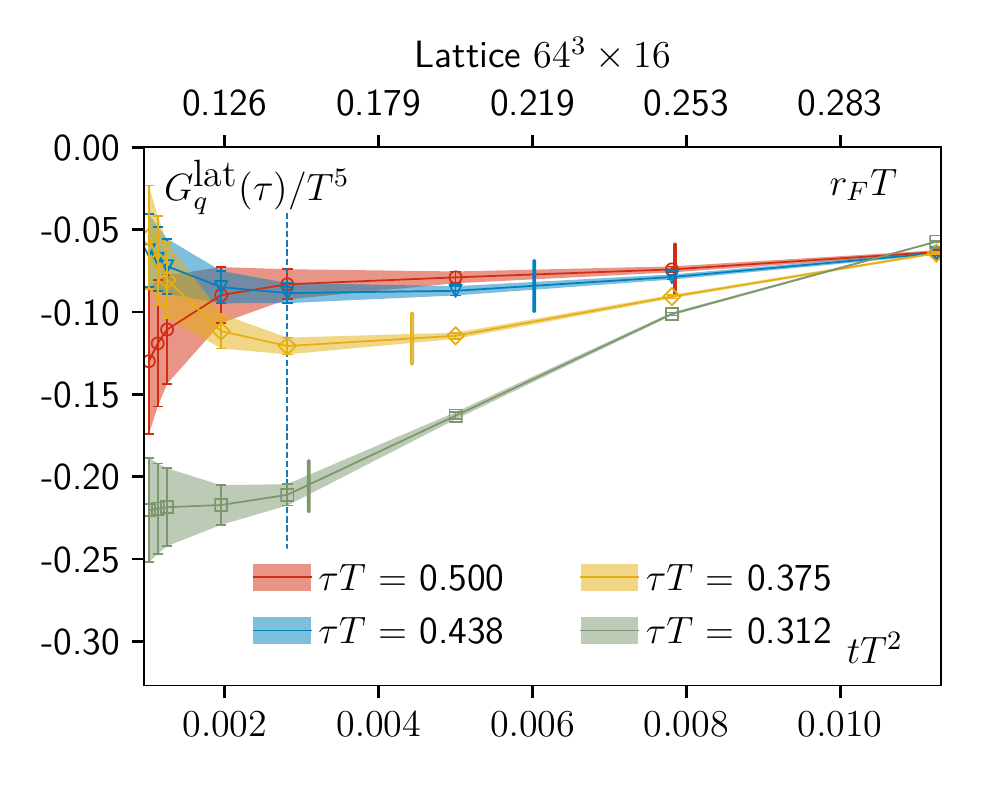}
    \vspace{-0.6cm}

    \raggedleft
    \includegraphics[width=1\linewidth]{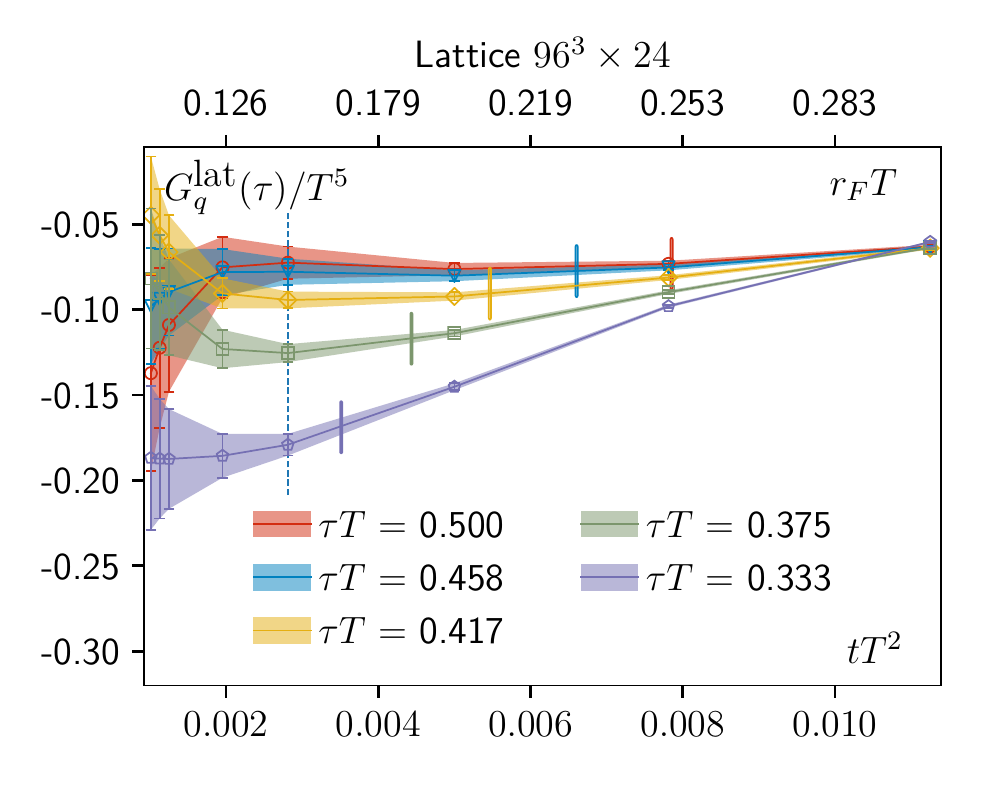}
\end{minipage}
\begin{minipage}{0.50\textwidth}

    \raggedright
    \includegraphics[width=1\linewidth]{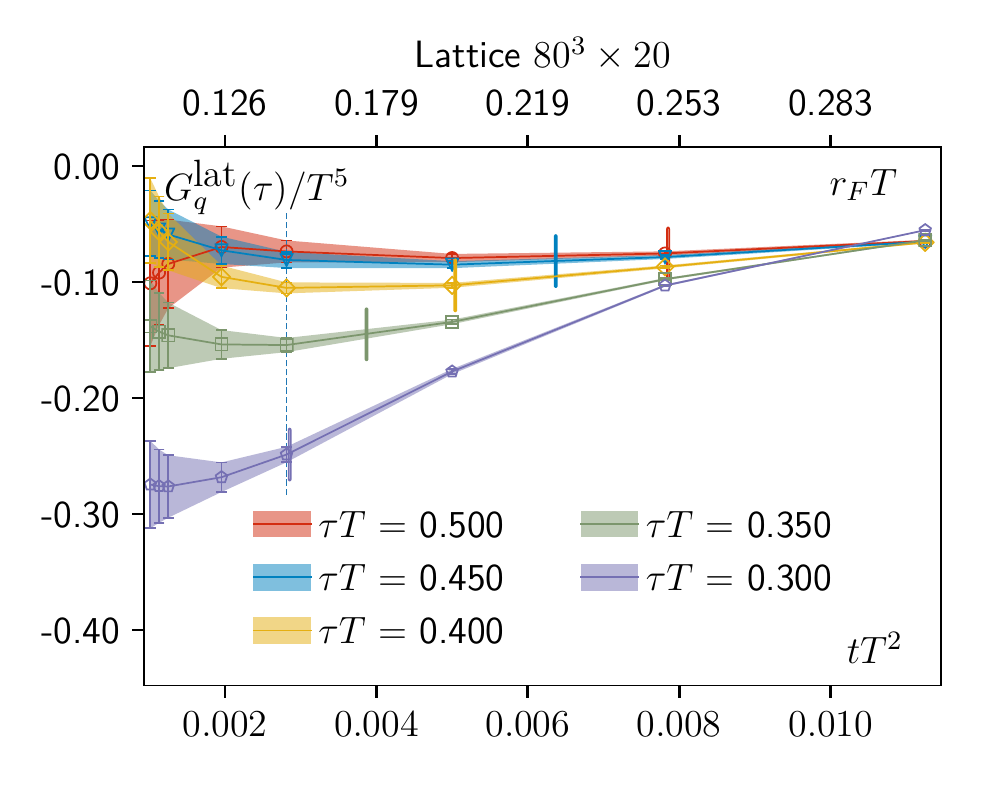}
    \vspace{-0.6cm}

    \raggedright
    \includegraphics[width=1\linewidth]{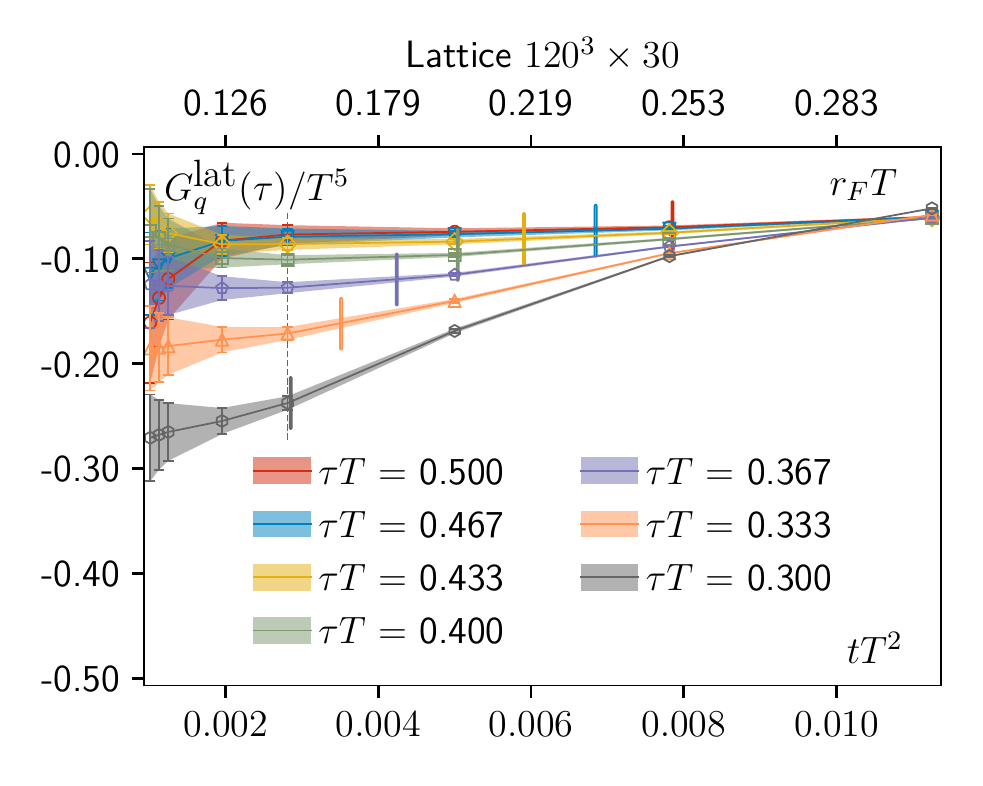}
\end{minipage}
    \caption{The correlation function $G^{\text{lat}}_{q}(\tau)$ with respect to the smoothing radius. Each curve is the correlation function at one particular separation. The vertical dotted line represents the lower limit $r_{F}T=0.15$, while the short vertical lines represent the limit \eqref{eq:flow_limit} of each curve.}
    \label{fig:corr_transposed}
\end{figure}

At separations $\tau T \ge 2r_{F}T$ it is not trivial how stable the correlation function upon flow is.
In order to investigate this, we plot the correlation function with respect to $\tau T^2$ in figure \ref{fig:corr_transposed}, where each curve is the correlation function at one particular separation. For small smoothing radii ($r_{F}T < 0.15$) UV-fluctuations are still visible, which is in agreement with the observations made for the topological susceptibility. 
For the largest separation, $\tau T=0.5$, we can see that it appears to have an almost $t$-independent plateau regime starting from $r_{F}T=0.15$ to $r_{F}T=0.25$. For $r_{F}T > 0.25$ the curves seem to become non-linear, which might be a consequence of the overlapping smoothing radii. However, to verify the last two observations we need to measure at more intermediate flow-times. 
Looking at the curves of the correlation function at lower separations we observe that the plateau regime decreases. 
This makes sense, because the correlation functions at lower separations are closer to the smoothing radius than at larger separations. Hence, a lower smoothing radius is enough to smoothen out the signal at these separations, and therefore, the lowest separations do not show any flow-time independent plateau regime.

\begin{figure}[t]

    \centering
\begin{minipage}{0.49\textwidth}
    \adjincludegraphics[width=1\textwidth,valign=t]{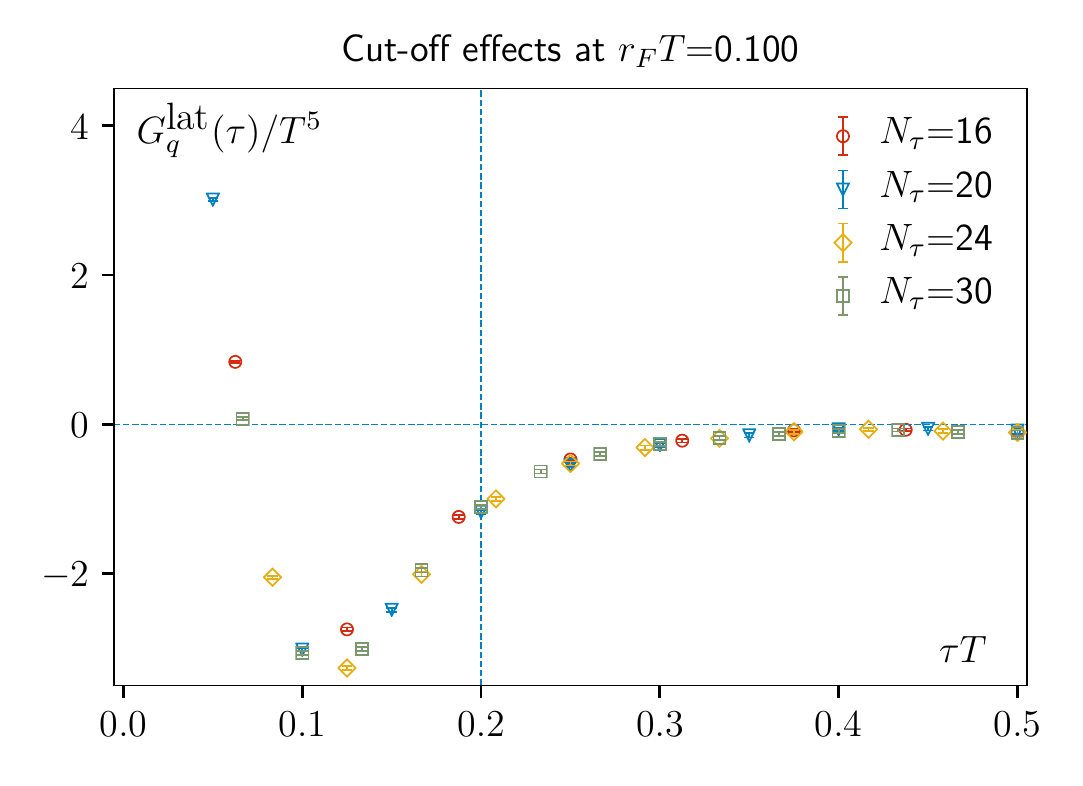}
\end{minipage}
\begin{minipage}{0.49\textwidth}
    \adjincludegraphics[width=1\textwidth,valign=t]{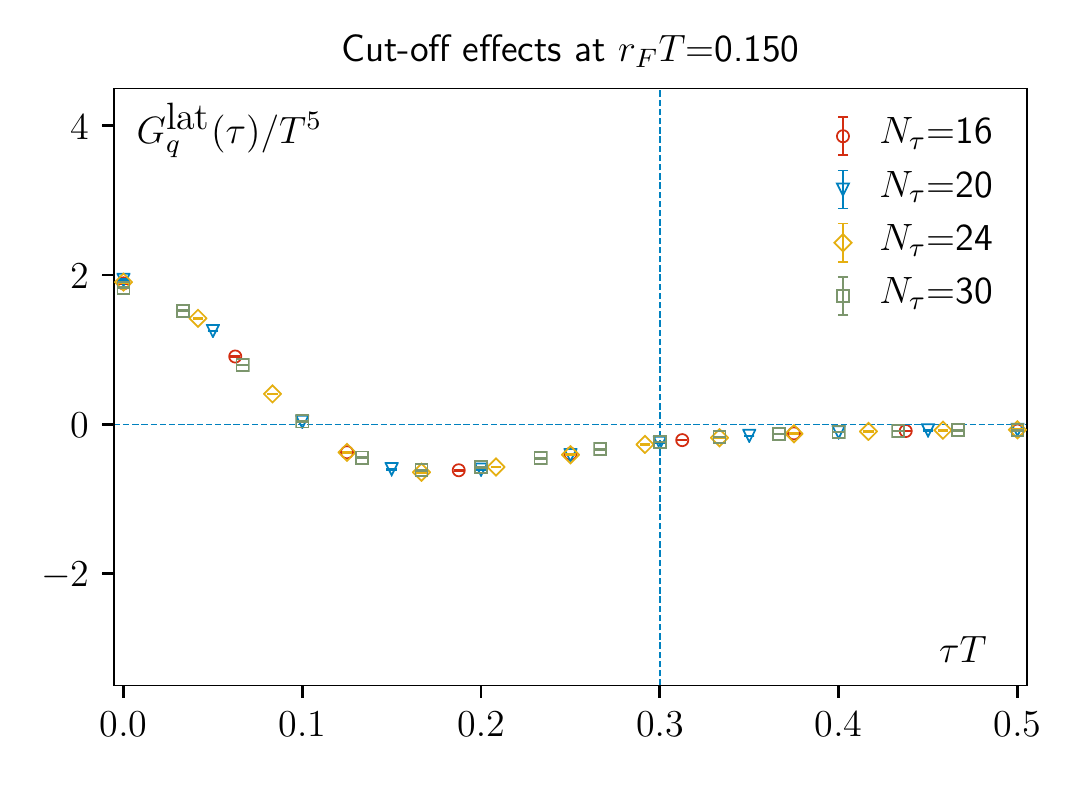}
\end{minipage}

    \caption{Correlation functions with respect to the separation at fixed flow time (left: $r_{F}T=0.100$, right: $r_{F}T=0.15$) and all discretizations. The vertical dotted line represents the limit $r_{F}T < \tau T/2$ (see eq. \eqref{eq:flow_limit})}
    \label{fig:corr_grad_cutOff}
\end{figure}

\begin{figure}[t]

    \centering
    \adjincludegraphics[width=.65\textwidth,valign=t]{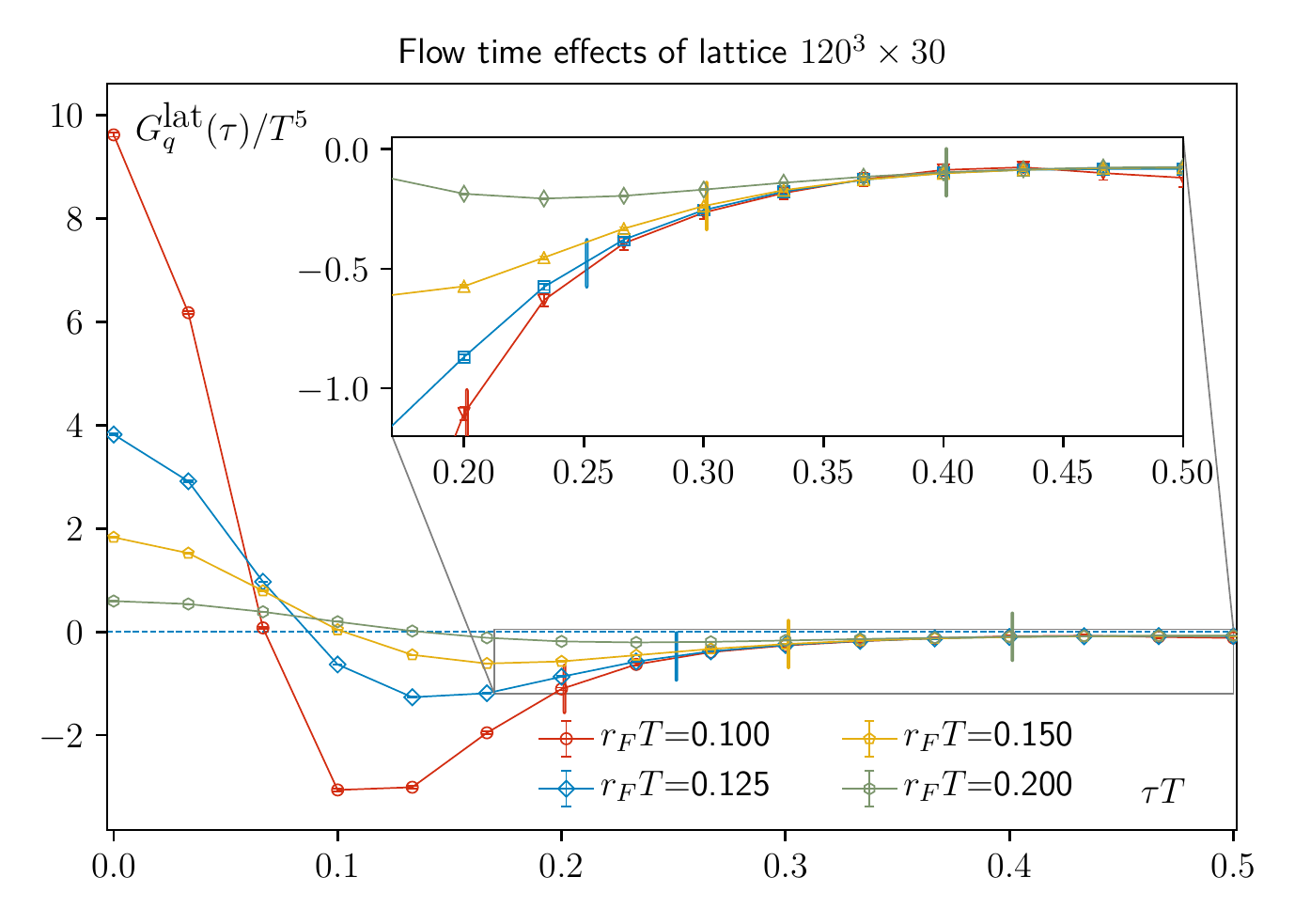}

    \caption{Correlation functions with respect to the separation of the $N_{\tau}=30$ lattice at different flow-times. The short vertical lines represent the limit $r_{F}T < \tau T/2$ (see eq. \eqref{eq:flow_limit}) of each curve, which indicates that only data points after these lines are valid.}
    \label{fig:corr_grad_flowEff}
\end{figure}

Figure \ref{fig:corr_grad_cutOff} shows the correlation function with respect to the separation at $r_{F}T=0.15$.
The correlation functions show only very small cut-off effects between the different lattices, which means that a continuum extrapolated curve wouldn't differ a lot from the lattice curves. This is expected, as the gradient flow shifts the gauge field closer to continuum and operators get renormalized. 
However, this would not be the final correlation function because, as mentioned in section \ref{sec:theory}, the correlation function in the continuum is expected to have a positive contact term only at $|\tau T|=0$, which is not seen in figure \ref{fig:corr_grad_cutOff}. If we have a look at the positive peak from $r_{F}T=0.15$ to $r_{F}T=0.1$, we observe that its width decreases as the flow-time decreases, which may indicate that the width collapses. So additionally, we need to perform a $t\rightarrow 0$ extrapolation after continuum extrapolating the correlation function. 

In Figure \ref{fig:corr_grad_flowEff} we show the correlation function of the $N_{\tau}=30$ lattice at different flow-times. We observe that the parts of the correlation function that fulfill the condition \eqref{eq:flow_limit} (i.e., the part after the vertical lines) lie almost on top of each other. 
If we put all pieces from this section together, we find that the correlation function on this setup seems to be flow-time independent if we take the condition \eqref{eq:flow_limit} into account and as long as the UV-fluctuations have been sufficiently smoothed out.

\section{Summary and Outlook}
Summarizing our results, we find that the gradient flow reveals a signal of the topological charge correlation function $G^{\text{lat}}_{q}(\tau)$ across lattices with different lattice spacings.
The correlation functions appear to have only small cut-off effects and smoothens out a large portion of the correlation function. Strictly speaking, all separations $ \tau T < 2r_{F}T$ are affected by flow effects. 
Therefore, we need finer lattices to obtain results at smaller distances.
The next step is to perform a continuum extrapolation and a flow-time extrapolation and compare the results with perturbation theory.
Furthermore, we will improve the error estimation by taking into account the autocorrelation times, induced by the freezing of topology on the fine lattices.
Our long-term goal is to extract the sphaleron rate from the correlation functions.

\acknowledgments
This work was supported by the Deutsche Forschungsgemeinschaft (DFG, German Research Foundation) - project number 315477589 - TTR 211; 
The numerical calculations were performed on the GPU cluster at Bielefeld University.

\bibliographystyle{JHEP}
\bibliography{bibliography}

\end{document}